\documentclass[runningheads]{llncs}

\usepackage{eccv}

\usepackage{eccvabbrv}

\usepackage{graphicx}
\usepackage{booktabs}
\usepackage{multirow}
\usepackage{makecell}
\usepackage{wrapfig}

\usepackage[accsupp]{axessibility}  

\usepackage{hyperref}

\usepackage{orcidlink}

\begin{document}

\title{A Watermark-Conditioned Diffusion Model for IP Protection} 

\titlerunning{Watermark-Conditioned Diffusion Model}

\author{Rui Min\inst{1} \and
Sen Li\inst{1} \and
Hongyang Chen\inst{2}\and
Minhao Cheng\inst{3}}

\authorrunning{R.~Min et al.}

\institute{Hong Kong University of Science and Technology \and Zhejiang Lab \and Pennsylvania State University\\
\email{\{rminaa,slien\}@connect.ust.hk,} \\ 
\email{hongyang@zhejianglab.com,} \email{mmc7149@psu.edu}
}

\maketitle

\begin{abstract}
  The ethical need to protect AI-generated content has been a significant concern in recent years. While existing watermarking strategies have demonstrated success in detecting synthetic content (\textbf{detection}), there has been limited exploration in identifying the users responsible for generating these outputs from a single model (\textbf{owner identification}). In this paper, we focus on both practical scenarios and propose a unified watermarking framework for content copyright protection within the context of diffusion models. Specifically, we consider two parties: the model provider, who grants public access to a diffusion model via an API, and the users, who can solely query the model API and generate images in a black-box manner. Our task is to embed hidden information into the generated contents, which facilitates further detection and owner identification. To tackle this challenge, we propose a \textbf{Wa}termark-conditioned \textbf{Diff}usion model called WaDiff, which manipulates the watermark as a conditioned input and incorporates fingerprinting into the generation process. All the generative outputs from our WaDiff carry user-specific information, which can be recovered by an image extractor and further facilitate forensic identification. Extensive experiments are conducted on two popular diffusion models, and we demonstrate that our method is effective and robust in both the detection and owner identification tasks. Meanwhile, our watermarking framework only exerts a negligible impact on the original generation and is more stealthy and efficient in comparison to existing watermarking strategies. Our code is publicly available at \url{https://github.com/rmin2000/WaDiff}.
\end{abstract}

\section{Introduction}
\label{sec:intro}
The recent progress in diffusion models has significantly advanced the field of AI-generated content (AIGC). Notably, several popular public APIs, such as Stable Diffusion~\cite{rombach2022high} and DALL$\cdot$E 3~\cite{betker2023improving}, have emerged, providing users with convenient access to create and personalize high-quality images. However, as these systems become more pervasive, the risk
of malicious use and attacks increases. In particular, some users with malicious intent may exploit the powerful generation capabilities of these models to create photo-realistic images like deep fakes~\cite{nightingale2022ai}, which can then be disseminated for illegal purposes. 
Moreover, as generative
models are excellent tools for creating and manipulating content, it is crucial to safeguard users’
copyrights and intellectual property when using state-of-the-art generative models. By discerning
the source of each user’s generated output, we can ensure that legitimate users’ contributions
are protected and prevent unauthorized replication or use of their content.
\begin{figure}
\centering

\includegraphics[width=9cm,height = 5.7cm]{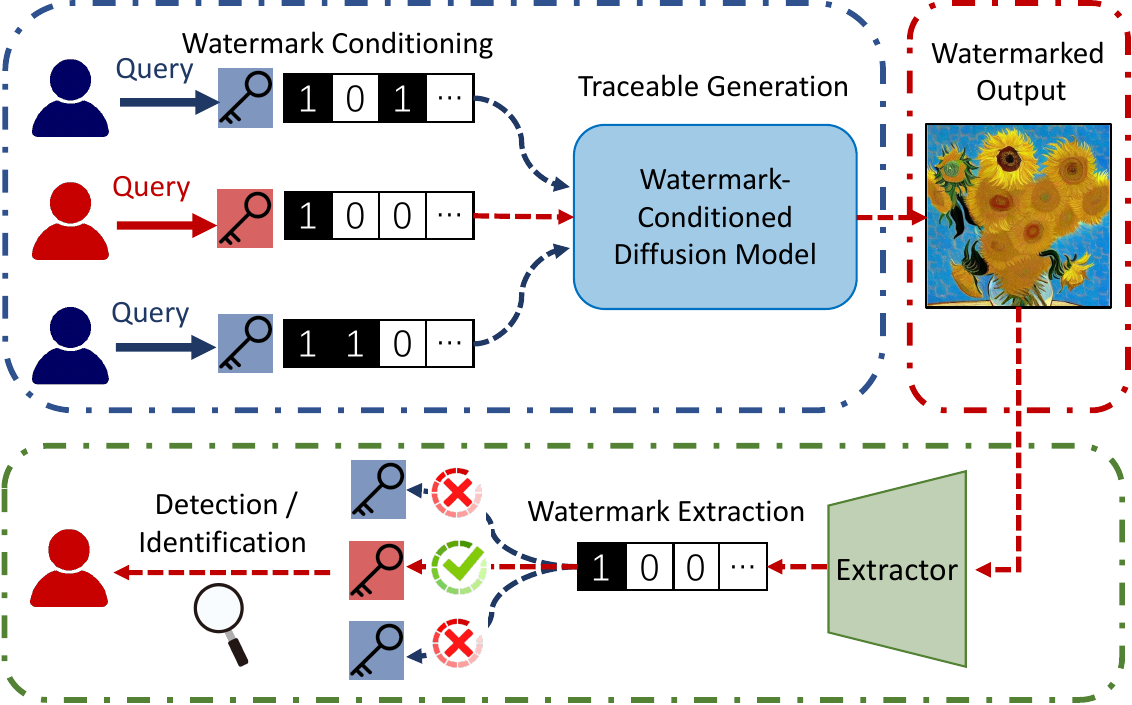}
\caption{Illustration of our proposed WaDiff. All users access the diffusion model by querying the public API and are assigned a unique watermark. The generation process is conditioned on the watermark, and each user's generated outputs would carry specific fingerprinting information which is further utilized to identify the owner of the generated image.}

\label{pic1}

\end{figure}

To enable the traceability of diffusion-generated images, a commonly employed strategy is to embed a unique fingerprint to contents generated by an individual user and then forensically identify the owner from the watermarked image. A line of previous works~\cite{barni1998dct,nikolaidis1998robust,ganic2004robust,singh2012novel,zhu2018hidden} has extensively investigated this approach within the realm of traditional multimedia copyright protection, which is commonly referred to as post-hoc watermark. Typically, these strategies involve embedding an imperceptible fingerprint into the generative content while leaving an identifiable trace that can be detected using a pre-designed mechanism. However, the post-hoc watermark requires additional computational costs for watermark injection and is more susceptible to circumvention. For instance, in the event of model leakage, attackers can easily detect and bypass the post-processing module. Recently, the Stable Signature~\cite{fernandez2023stable} investigated the latent diffusion paradigm and proposed a method that incorporates watermarks into each latent decoder. However, solely fingerprinting the latent decoder limits their application scenario to the latent diffusion paradigm only and can be easily circumvented by retraining the latent decoder on a clean dataset. At the same time, since only a fixed watermark could be embedded into the latent decoder, every model has to be fine-tuned before being distributed to the users, making it hard to use in large-scale real-world systems. Considering both the scalability and effectiveness, we aim to investigate whether we can embed fingerprints during the generation process to incorporate user-specific watermarks without customized fine-tuning.

In this paper, we introduce WaDiff (as shown in Figure~\ref{pic1}), a watermark-conditioned diffusion model, which incorporates the watermark as a conditioned input and generates images with unique fingerprints tailored to individual users. Unlike previous approaches~\cite{fernandez2023stable,xiong2023flexible}, which solely focus on fingerprinting the latent decoder in the diffusion model, our watermarking strategy is seamlessly integrated into the diffusion generation process. This makes our approach more general and applicable to other types of diffusion models~\cite{ho2020denoising} that do not include a latent decoder, while simultaneously eliminating the need for post-processing. Our WaDiff builds upon a pre-trained diffusion model, with slight modifications to the original input layers to accommodate the inclusion of watermark information by expanding the channels. Specifically, we first embed the watermark bits through a linear layer and then concatenate this projected vector with the original input to construct the watermark-conditioned input. We design a unified watermarking framework with two novel objective functions, named message retrieval loss and consistency loss. The message retrieval loss ensures the effective embedding of fingerprints into the generated content, allowing for successful retrieval of the embedded watermark. Meanwhile, the consistency loss ensures that the inclusion of the watermark has a negligible effect on the overall generation quality.

To this end, we evaluate our method on two popular open-source diffusion models and perform both detection and identification tasks. We also make detailed comparisons with widely used post-hoc watermarking strategies and Tree-Ring \cite{wen2023tree}, a training-free fingerprinting framework for diffusion models. Experimental results demonstrate that our efficient watermarking strategy enables accurate detection and identification in a large-scale system with numerous users and remains robust across various data augmentations. Moreover, the generated images after watermarking maintain exceptional generation quality and are visually indistinguishable across different watermarks. Our contributions are organized as follows:

\begin{itemize}
    \item Different from previous work, we propose a scalable watermarking strategy that efficiently integrates user-specific fingerprints into the diffusion generation process without the need for customized fine-tuning.
   
    \item We introduce WaDiff, a watermark-conditioned diffusion model, along with a unified watermarking framework. Unlike post-hoc fingerprinting, WaDiff manipulates the watermark as conditioned input and allows for effective watermarking while having a negligible impact on the generation quality.
    
    \item Extensive experiments demonstrate that our watermarking framework achieves precise and robust performance in detecting AI-generated content and identifying the source owner of generated images.
\end{itemize}

\section{Related Work}
\label{sec:formatting}

\subsection{Diffusion Models}
Diffusion models have recently shown tremendous capability for high-quality image generation~\cite{ho2020denoising, song2020denoising, saharia2022photorealistic, rombach2022high, song2020score}. Depending on the space in which the generation process is performed, diffusion models can be divided into pixel-space~\cite{ho2020denoising, dhariwal2021diffusion} and latent-space diffusion models~\cite{rombach2022high}. In pixel-space diffusion models, images are directly generated starting from sampled Gaussian noise~\cite{ho2020denoising}. To reduce the computational complexity, latent diffusion models have been proposed to first generate latent features from noise and then decode the latent features to images by VAE~\cite{rombach2022high}. With the powerful generation capability, they have demonstrated amazing results on various computer vision tasks, such as text-to-image generation~\cite{nichol2021glide,saharia2022photorealistic}, sketch-to-image generation~\cite{meng2021sdedit}, text-guided image editing~\cite{nichol2021glide}.

\subsection{Image Watermarking}
Image watermarking has been a broadly explored technique for decades. Traditional strategies typically start from a host image and inject watermark information directly in the spatial domain~\cite{nikolaidis1998robust, singh2012novel}, or through certain domain transformations such as Discrete Cosine Transform (DCT)~\cite{barni1998dct} and Discrete Wavelet Transform (DWT)~\cite{ganic2004robust}. With advancements in deep learning, researchers have also explored the use of deep-learning techniques to replace manually designed watermark patterns. For example, a line of works~\cite{zhu2018hidden, liu2019novel,tancik2020stegastamp,zhong2020automated,zhang2020udh} leverages the capability of Deep Neural Networks (DNNs) to improve the stealthiness and robustness of the watermark. In addition to facilitating image protection, image watermarking techniques have been proved useful in various security challenges, such as model ownership verification~\cite{zhang2020model, wu2020watermarking, zhang2018protecting}, backdoor attack~\cite{li2021invisible}, dataset copyright protection~\cite{li2022untargeted}, and forensic adversarial defense~\cite{cheng2023identification}. In our method, we leverage image watermarking to embed unique watermark information for individual users, thereby facilitating subsequent detection and identification.

\subsection{Fingerprinting in Diffusion Models}
In light of recent advancements in generative AI, researchers have been exploring watermarks to safeguard or regulate the usage of generative models~\cite{yu2020responsible,yu2021artificial,kirchenbauer2023watermark,zhao2023recipe,cui2023diffusionshield,ma2023generative,fernandez2023three, ci2024ringid, yang2024gaussian}. In this study, we primarily focus on fingerprinting the generative content. The Stable Signature~\cite{fernandez2023stable} first proposed a watermarking scheme for latent diffusion models~\cite{rombach2022high} by fingerprinting a set of latent decoders. They then distributed these customized decoders to individual users for both detection and identification. Building upon this work,~\cite{xiong2023flexible} improved the scalability by incorporating a message matrix into the latent decoder, allowing multiple watermarks to be carried within a single architecture. However, their methods only focus on manipulating the latent decoder while keeping the diffusion process intact, which makes their methods only applicable to latent-space diffusion models. Also, this leaves them vulnerable to attacks like simply retraining the fingerprinted latent decoder on a clean dataset. In a parallel work called Tree-Ring~\cite{wen2023tree}, researchers concealed the fingerprint within the frequency domain of the initial noisy vector. Detection is then performed by reversing the generative image back to the noisy vector and comparing it with the original noisy pattern. Despite its training-free nature, the detection process heavily relies on the time-consuming reversion process and poses challenges for identification among multiple users.

In contrast to previous studies, we propose the WaDiff along with a simple and unified watermarking framework for both detection and owner identification. Our method seamlessly integrates the fingerprinting process into the image generation process, which can be applied with various diffusion types and sampling schedules, resulting in enhanced stealthiness and compatibility.

\section{Problem Setting}
The model provider deploys a generation model and grants public access to $m$ registered users. Considering the intellectual property (IP) protection, the model architecture and parameters are concealed from users, which means only black-box access for users is permitted and all the internal information remains encrypted. Note that the black-box scenario is common in practice, and has been widely adopted in current generative models such as the Midjourney and ChatGPT~\cite{brown2020language}. Each user denoted as $u_i$, is required to register before usage and would be then assigned a unique ID $i$ where $i \in \{1,\dots,m\}$, representing the user's unique identity. One of the users $u_{i}$ would use the provided generation model to generate a picture $\mathbf{p}$. In practice, the $\mathbf{p}$ might be further processed under several image manipulations $f$ such as resizing and compression. We aim to build a security auditing system to enable the detection and tracing of the diffusion-generated content $\mathbf{p}$. 

We divided our task into two challenges. The first challenge involves determining whether the generative content originates from our diffusion model, which we refer to as detection. However, simple detection of the generative content is insufficient to differentiate who generated a particular image when multiple users are engaged. Therefore, our second challenge focuses on a more complex scenario, which entails accurate identification of the owner of $f(\mathbf{p})$, from a pool of users. To address both challenges effectively, we propose a unified watermarking framework named WaDiff, which incorporates the embedding of user-specific fingerprints within the generation process. In other words, the generative images of each user are watermarked with unique information that distinguishes them from other users and facilitates both detection and identification.

\section{Methodology}
\label{sec:4}
\subsection{Preliminaries of Diffusion Models}
\label{sec:4.1}
In this section, we briefly introduce the notation for a vanilla DDPM. The basic diffusion model typically involves two critical components known as the forward and backward processes. The forward process is a Markov chain that gradually adds noise into a real data sample $\mathbf{x}_0 \sim q(\mathbf{x}_0)$ over $T$ steps. Specifically, for each time step $t \in \{1,\dots,T\}$, the latent $\mathbf{x}_t$ is obtained by adding Gaussian noise to the previous latent $\mathbf{x}_{t-1}$. Alternately, the noising procedure can be viewed as sampling from the distribution $q(\mathbf{x}_t|\mathbf{x}_{t-1})=\mathcal{N}(\sqrt{1-\beta_t}\mathbf{x}_{t-1}, \beta_t\mathbf{I})$, which could be re-parameterized as:
\begin{equation}
    \label{eq1}
    \mathbf{x}_t = \sqrt{1-\beta_t}\mathbf{x}_{t-1} + \sqrt{\beta_t}\epsilon,
\end{equation}
where $\epsilon \in \mathcal{N}(\mathbf{0}, \mathbf{I})$ is a Gaussian noise, and $\beta_t$ is a predefined time-dependent variance schedule. By substituting $\alpha_t=1-\beta_t$ and $\overline{\alpha}_t=\prod_{s=1}^{t}{\alpha_s}$, the Equation \ref{eq1} could be further simplified to its closed form:
\begin{equation}
\label{eq2}
    \mathbf{x}_t = \sqrt{\overline{\alpha}_t}\mathbf{x}_0 + \sqrt{1-\overline{\alpha}_t}\epsilon.
\end{equation}
Contrary to the forward process, the backward process aims to gradually reverse each forward step by estimating the latent $\mathbf{x}_{t-1}$ from $\mathbf{x}_{t}$. Starting from the $\mathbf{x}_T$, the diffusion model parameterized by $\theta$, predicts the learned estimate of the Gaussian noise in Equation \ref{eq2} as $\epsilon_{\theta}(\mathbf{x}_T)$ and then the $\mathbf{x}_{T-1}$ could be sampled from the distribution denoted as $p(\mathbf{x}_{T-1}|\mathbf{x}_{T}, \epsilon_{\theta}(\mathbf{x}_T))$. In the subsequent denoising procedure, a similar recovery from $\mathbf{x}_{t}$ to $\mathbf{x}_{t-1}$ is repeated until the restoration of the original input $\mathbf{x}_0$. 

\subsection{Rooting Fingerprints in Diffusion Process}

\paragraph{\textbf{Pre-training Watermark Decoders.}}
Inspired by~\cite{fernandez2023stable}, we first train an image encoder $\mathcal{W}$ and decoder $\mathcal{D}$ such that $\mathcal{D}$ could retrieve the message $\mathbf{w}$ pre-embedded by $\mathcal{W}$. 
The $\mathcal{W}$ is then discarded and only the pre-trained $\mathcal{D}$ is left to serve as a reference to fine-tune the diffusion model. 
Specifically, the encoder $\mathcal{W}$ takes image $\mathbf{x}$ and $n$-bit binary message $\mathbf{w} \in \{0,1\}^{n}$ as input and encodes $\mathbf{w}$ as an imperceptible residual $\delta$ added to $\mathbf{x}$, where the extractor $\mathcal{D}$ aims to restore the pre-encoded $\mathbf{w}$ from the watermarked input $\mathcal{W}(\mathbf{x})=\mathbf{x}+\delta$. To train $\mathcal{W}$ and $\mathcal{D}$ in an end-to-end manner, we utilize two loss terms to fulfill the optimization objective. First, to accurately restore $\mathbf{w}$ from $\mathcal{W}(\mathbf{x})$, we minimize the Binary Cross Entropy (BCE) between the decoded output $\mathcal{D}(\mathcal{W}(\mathbf{x}))$ and ground-truth $\mathbf{w}$. Second, to reduce the visibility of $\delta$, we penalize $\|\delta\|_2$ such that the added message perturbation is less perceptible. We further notice that the $\mathcal{D}$ obtained with the above training schedule is sensitive to several data manipulations such as resizing and compression. Therefore, similar to the simulation layer in~\cite{fernandez2023stable}, we transform $\mathcal{W}(\mathbf{x})$ with random data augmentations during the training stage to improve the robustness of the watermark system. We formulate the complete training objective as follows:
\begin{equation}
    \min_{\mathcal{W}, \mathcal{D}} \mathbb{E}_{\mathbf{x}, \mathbf{w},f}[\mathcal{L}_{BCE}(\mathcal{D}(f(\mathcal{W}(\mathbf{x}))),\mathbf{w}) + \gamma\|\delta\|_2],
\end{equation}
where $\gamma > 0$ is a hyperparameter to control the visibility of $\delta$ and $f$ is a randomly selected image transformation from a pool of data augmentations.


\paragraph{\textbf{Watermark-conditioned Diffusion Model.}}
In contrast to post-hoc strategies, where the fingerprinting process occurs after the entire generation process, our approach imprints fingerprints during the sampling process.
To discern the source user of generated outputs, we assign each user $u_i$ a unique $\mathbf{w}_i$ as the conditioned input of the diffusion model such that $\mathbf{w}_i$ could be embedded to the generated content during inference and correctly restored by the pre-trained $\mathcal{D}$.

To incorporate $\mathbf{w}_i$ into the generation process, we expand the input channels and conceal the watermarking information at each denoising step $t$. This is achieved by applying a linear layer to project $\mathbf{w}_i$ into $\mathcal{P}(\mathbf{w}_i) \in \mathbb{R}^{\widetilde{C} \times H \times W}$, where $\widetilde{C}$ denotes the number of watermark channels in the latent space. We then concatenate it with the original latent variable $\mathbf{x}_t \in \mathbb{R}^{C \times H \times W}$ along their first dimension, resulting in the conditioned latent variable $\mathbf{\hat{x}}_{t, i} = concat(\mathbf{x}_t, \mathcal{P}(\mathbf{w}_i)) \in \mathbb{R}^{(C+\widetilde{C})\times H\times W}$. Note that channel expansion is a common manipulation to integrate additional information and has been widely used in previous works~\cite{zhu2018hidden,tancik2020stegastamp}. The next step involves embedding $\mathbf{w}_i$ into the generative content to allow for detection by the pre-trained decoder $\mathcal{D}$. However, as the sampling process typically involves multiple denoising steps to obtain the generative output, it is challenging to directly incorporate it into the fine-tuning process. To address this issue, we start by restoring the original image $\mathbf{x}_0$ within a single step. Revisiting the forward noising in Equation \ref{eq2}, $\mathbf{x}_0$ could be directly recovered from $\mathbf{x}_t$ by subtracting the second noise term and subsequently scaling. In the absence of the ground truth $\epsilon$, we take the prediction from the diffusion model as an estimate. In our watermark-conditioned model, we replace $\epsilon$ with $\epsilon_{\theta}(\hat{\mathbf{x}}_{t, i})$ and construct the conditioned reverse of $\mathbf{x}_0$ at time step $t$ as:
\begin{equation}
\label{eq4}
    \mathbf{x}_{0, i}^t = \frac{\mathbf{x}_t - \sqrt{1-\overline{\alpha}_t}\epsilon_{\theta}(\hat{\mathbf{x}}_{t, i})}{\sqrt{\overline{\alpha}_t}}.
\end{equation}
We then formulate our first optimization objective as optimizing the \textbf{message retrieval loss}, which we defined as follows:
\begin{equation}
\label{eq5}
    \min_{\theta} \mathbb{E}_{\mathbf{x}, i, t}[\mathcal{L}_{m}(\mathcal{D}(\mathbf{x}_{0, i}^t), \mathbf{w}_i)].
\end{equation}
By optimizing Equation \ref{eq5}, we ensure that the pre-embedded $\mathbf{w}_i$ in $\mathbf{x}_{0, i}^t$ could be detected by $\mathcal{D}$. However, the image quality of $\mathbf{x}_{0, i}^t$ is significantly influenced by the denoising step $t$, resulting in higher image quality for smaller $t$ and noisier images for larger $t$. Therefore, we empirically introduce a threshold $\tau$ where we only minimize the message retrieval loss when $t\leq\tau$. This strategy helps effectively inject the watermark while stabilizing the fine-tuning procedure.

\begin{figure*}
\centering
\includegraphics[width=12cm, height=4cm]{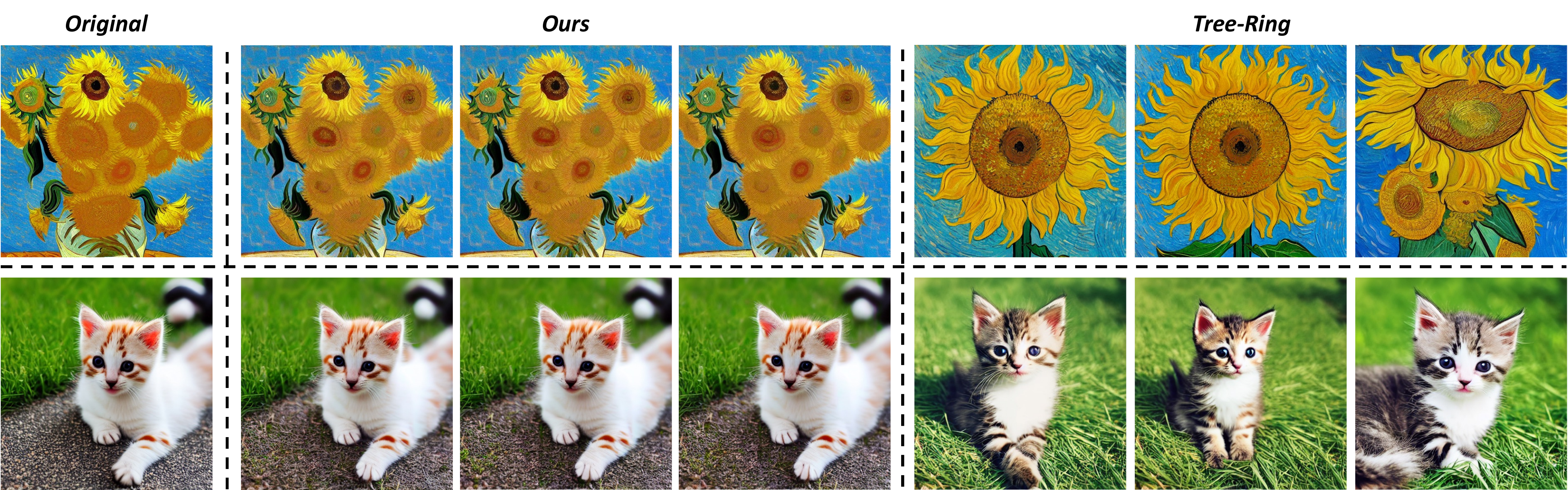}
\caption{Watermarked examples of our method and $\text{Tree-Ring}_{Rings}$ sampled from the Stable Diffusion. It is observed that our method achieves a substantial improvement in image consistency among images with diverse watermarks.}
\label{fig:2}
\end{figure*}

\paragraph{\textbf{Preserving Image Consistency.}}
In addition to fingerprinting the output, it is critical to preserve the generated images' quality after watermarking. In other words, the generated images after watermarking across different users should be visually equivalent. To achieve this, we treat the original model as an oracle and align the generated output with the original output. Since the whole generation involves multiple denoising steps, we need to ensure the $\epsilon_{\theta}(\hat{\mathbf{x}}_{t, i})$ and $\epsilon_{\theta_{ori}}(\mathbf{x}_t)$ are aligned for each $t$, where $\theta_{ori}$ represents the pre-trained diffusion model weights. Therefore, we introduce the second optimization objective by minimizing over the image \textbf{consistency loss}, denoted as follows:
\begin{equation}
\label{eq6}
    \min_{\theta} \mathbb{E}_{\mathbf{x}, i, t}[\mathcal{L}_c(\epsilon_{\theta}(\hat{\mathbf{x}}_{t, i}), \epsilon_{\theta_{ori}}(\mathbf{x}_t))],
\end{equation} where $\mathcal{L}_c$ is the Mean-Squared Loss. To further improve the image consistency between different watermarks, for $t>\tau$, we replace the conditioned input $\mathbf{w}_i$ with a never-used null watermark denoted as $\mathbf{w}_{null}$. By fixing the condition with $\mathbf{w}_{null}$, we ensure that $\epsilon_{\theta}(\hat{\mathbf{x}}_{t, null})$ is distinct from $\mathbf{w}_i$ and keeps unchanged until $t \leq \tau$, which yields an improved image consistency among diverse users. We demonstrate watermarked samples of our method along with Tree-Ring~\cite{wen2023tree} in Figure~\ref{fig:2} and can observe that our method significantly improves the image consistency between watermarked contents. From the results, our watermarked outputs are not only visually equivalent across different users but also maintain the original semantic meaning with only a slight visual difference.

\begin{figure*}
\centering

\includegraphics[width=11cm, height = 7.5cm]{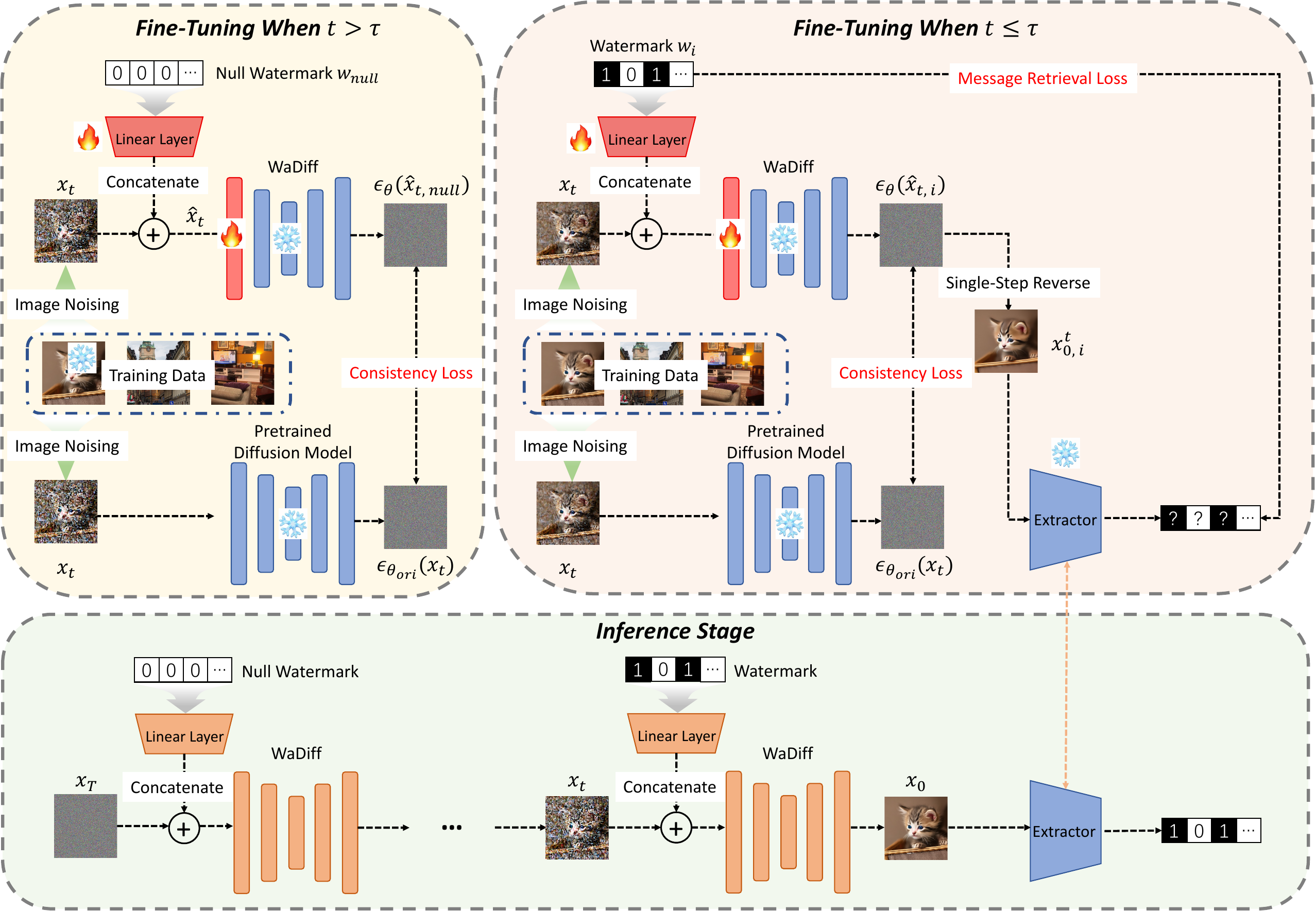}
\caption{Overview of our proposed watermarking framework. The top two figures illustrate our fine-tuning process. For $t>\tau$, we solely focus on preserving image consistency and incorporate a null watermark. For $t\leq \tau$, we integrate the normal watermark and introduce an additional message retrieval loss to embed watermarks. The inference stage is depicted below, where we inject the null watermark when $t>\tau$ and transition it to the payload watermark when $t\leq\tau$.}
\label{fig:framework}
\end{figure*}

\paragraph{\textbf{End-to-end Fine-tuning.}}
Instead of fine-tuning the entire diffusion model, we selectively fine-tune the watermark projector $\mathcal{P}$ and the first input block while keeping the remaining weights unchanged. We adopt this approach for two reasons. First, we observe that fine-tuning the first block is sufficient for effective watermark injection. Our method achieves comparable performance to fine-tuning the entire architecture but with faster speed and lower memory cost. Additionally, we have empirically observed that fine-tuning the entire model can lead to unstable generation, resulting in watermarked images with significantly compromised generative quality. A detailed analysis of fine-tuning various model subsections is provided in the \textit{Appendix}. Formally, we separate the whole parameters $\theta$ into two sets: $\theta_{head}$, representing the parameters of $\mathcal{P}$ and the first input block, and $\theta_{tail}$, representing the remaining weights, \textit{i.e.}, $\theta=\{\theta_{head},\theta_{tail}\}$. During fine-tuning, only $\theta_{head}$ is optimized while $\theta_{tail}$ is fixed. In sum, we incorporate both optimization objectives above and formulate the fine-tuning process as:
\begin{equation}
\begin{aligned}
    \min_{\theta_{head}} \mathbb{E}_{\mathbf{x}, i, t}[\mathbb{I}(t\leq\tau)(\mathcal{L}_c(\epsilon_{\theta}(\hat{\mathbf{x}}_{t, i}), \epsilon_{\theta_{ori}}(\mathbf{x}_t)) + \eta\mathcal{L}_{m}(\mathcal{D}(\mathbf{x}_{0, i}^t), \mathbf{w}_i)) \\
    + \mathbb{I}(t>\tau)\mathcal{L}_c(\epsilon_{\theta}(\hat{\mathbf{x}}_{t, null}), \epsilon_{\theta_{ori}}(\mathbf{x}_t))],
\end{aligned}
\end{equation}
where $\mathbb{I}$ represents the indicator function and $\eta$ controls the trade-off between image consistency and watermarking effectiveness. Our framework is demonstrated in Figure \ref{fig:framework}.

\subsection{Detection and Identification}
Once the diffusion model is fingerprinted, the generated content for each $u_i$ will contain specific information that is conditioned by $\mathbf{w}_i$. Given a candidate $\mathbf{p}$, we can then recover the source watermark $\mathbf{w}_s=\mathcal{D}(\mathbf{p})$ using the pre-trained $\mathcal{D}$ model. Once we have recovered $\mathbf{w}_s$, we can utilize it for both detection and identification. To detect whether $\mathbf{p}$ belongs to our model, we set a bit threshold $\tau_b$ and calculate the number of matched bits $M_{i}=\mathbf{w}_s\odot \mathbf{w}_i$, where $\odot$ represents the XNOR function. If $\frac{M_{i}}{n}>\tau_b$, we will conclude that $\mathbf{p}$ is generated from our model; otherwise, it is generated from other sources. For identification among $m$ users, we can determine the owner by finding the corresponding user ID whose watermark $\mathbf{w}_i$ has the best match with $\mathbf{w}_s$, which can be formulated as follows:
\begin{equation}
    \arg\max_{i} M_{i},\ \ i\in\{1\dots m\}.
\end{equation}

\section{Experiments}
To evaluate the identification performance of our proposed WaDiff, we implement our method on two widely used diffusion models: a text-to-image latent diffusion model and an unconditional diffusion model. We conduct comprehensive comparisons with other watermarking strategies including a traditional strategy DwtDct~\cite{cox2007digital}, a deep-learning-based steganography technique StegaStamp~\cite{tancik2020stegastamp}, which we employed in our pre-training stage, and a recently proposed training-free framework Tree-Ring~\cite{wen2023tree}. Note that we have not included the Stable Signature~\cite{fernandez2023stable} in our comparison since it considered the model distribution setting which is different from ours. Besides, it necessitates re-training diverse latent decoders, which is not scalable in our identification task. We also conduct thorough ablation studies and provide detailed analysis in the following sections.

\subsection{Experimental Settings}
\paragraph{\textbf{Model and Dataset.}} For the text-to-image latent diffusion model, we utilize the Stable Diffusion V1.4 as our pre-trained model, and for the unconditional diffusion model, we utilize the 256 $\times$ 256 ImageNet diffusion model. To fine-tune the diffusion models, we randomly select 5000 images with corresponding text descriptions from the training set of MS-COCO 2014~\cite{lin2014microsoft} for the Stable Diffusion and 5000 images from the training set of ImageNet~\cite{deng2009imagenet} for the 256x256 ImageNet diffusion model.

To benchmark the effectiveness of our proposed strategy, we employ different metrics for detection and identification evaluation. For detection, we report the area under the curve (AUC) of the receiver operating characteristic (ROC) curve. For identification, we begin by generating a pool of $m$ users, each associated with a unique binary code. Randomly drawn from this pool, we select a subset of users and generate images for each selected user. The identification performance of our method is evaluated using the tracing accuracy metric defined as:
\begin{equation}
\text{Trace Acc} = \frac{N_{\text{correct}}}{N_{\text{total}}},
\end{equation}
where $N_{\text{correct}}$ represents the number of correctly identified images and $N_{\text{total}}$ is the total number of candidate images for identification. Additionally, we assess the image consistency by calculating the structural similarity index (SSIM)~\cite{wang2004image} between pairs of watermarked content. Furthermore, we measure the impact of watermarking on the original generation quality by reporting the difference between the Frechet Inception Distance (FID)~\cite{heusel2017gans} of the watermarked contents and the originally generated contents.

\paragraph{\textbf{Implementation Details.}}
All experiments were conducted utilizing 8 NVIDIA A100 GPUs. The fine-tuning of both diffusion models involved employing an AdamW optimizer~\cite{loshchilov2017decoupled} with a learning rate of $1e^{-4}$. For the Stable Diffusion model, we set $\tau$ to 500 and $\eta$ to 0.05, while for the ImageNet diffusion model, we selected $\tau$ as 400 and $\eta$ as 0.25. Notably, in addition to directly aligning the predicted noise of the ImageNet diffusion model with the original model, we empirically found that aligning the single-step-reverse images resulted in improved image quality. Implementation details and additional hyperparameter evaluations are deferred in the \textit{Appendix}. We fine-tune 40 epochs for the Stable Diffusion model and 25 epochs for the ImageNet diffusion model. For the Stable Diffusion model, we set the default guidance scale to 7.5 and used text descriptions from the validation set of MS-COCO 2014~\cite{lin2014microsoft} as prompts. We adopt the watermark length to 48 by default, which is commonly used in previous work~\cite{fernandez2023stable}, and provide experiments of other lengths in the \textit{Appendix}. We adopted the DDIM~\cite{song2020denoising} sampler with 50 sampling steps as default for both models. More implementation details on training baseline methods and the watermark decoder pre-training is in the \textit{Appendix}.

\begin{table*}[htbp]
\caption{This table includes our main results.  $\text{Trace}\ m$ indicates the tracing accuracy (\%) of our identification among $m$ users in total.}
\label{tab:main_exp}
\setlength\tabcolsep{1pt}
\centering
    \begin{sc}
\resizebox{0.95\textwidth}{!}{\begin{tabular}{c|c|cccccccc}
\toprule
Model      &Type       & Method   &AUC  &  $\text{Trace}\ 10^4$ & $\text{Trace}\ 10^5$ & $\text{Trace}\ 10^6$  &Trace Avg & SSIM$(\uparrow)$ &FID Diff$(\downarrow)$ \\ \midrule
\multirow{4}{*}{\makecell[c]{Stable \\ Diffusion}} & \multirow{2}{*}{\makecell[c]{Post \\ Generation}} 
                    
                    & DwtDct  &0.917& 76.30  & 74.70 & 72.90 &74.63  &0.999&-0.36\\ 
                    && StegaStamp & 1.000  & 99.98 & 99.98 & 99.96 &99.97 &0.999&+0.27\\ \cmidrule{2-10} 
                    &\multirow{3}{*}{\makecell[c]{Merged \\ Generation}}& $\text{Tree-Ring}_{Rand}$   &0.999&0.04&0.00  &0.00 &0.01  & 0.457 & +0.14 \\
                    && $\text{Tree-Ring}_{Rings}$  &0.999& 0.00  & 0.00 & 0.00  & 0.00 & 0.575 & +0.77\\ 
                    && WaDiff (Ours)   &0.999& 98.20 & 96.76 & 93.44 &96.13&  0.999 &+0.41\\ \midrule
\multirow{4}{*}{\makecell[c]{256$\times$256 \\ ImageNet}} & \multirow{2}{*}{\makecell[c]{Post \\ Generation}} & DwtDct&0.936 & 71.30  & 68.10 & 65.20&68.20  &0.997&-0.05 \\ 
                    && StegaStamp &1.000& 99.98  & 99.98 & 99.98 &99.98 &0.998&+0.11\\ \cmidrule{2-10} 
                    &\multirow{3}{*}{\makecell[c]{Merged \\ Generation}}& $\text{Tree-Ring}_{Rand}$ &0.999 &0.00  &0.00  &0.00  &0.00  & 0.584 & +0.17 \\ 
                    && $\text{Tree-Ring}_{Rings}$   &0.999& 0.00  & 0.00 &0.00 & 0.00& 0.652 & +0.23 \\
                    && WaDiff (Ours)  & 1.000 &99.68  &99.38  &98.78  &99.28 &0.997&+0.08\\ \bottomrule
\end{tabular}
}
 \end{sc}

\end{table*}

\subsection{Detection and Identification Results}
In this section, we present the main experimental results of our method. For the detection task, we use 5000 watermarked images along with another 5000 clean images sampled from the original pre-trained diffusion model to calculate the AUC. As for owner identification, we evaluate our method using different sizes of user pools, ranging from ten thousand to one million users. For each user pool, we randomly select 1000 users and generate 5 images per user, resulting in a total of 5000 images. The tracing accuracy is then calculated based on these watermarked images. For SSIM calculation, we first randomly select 200 distinct initial noises and generate a group of 5 images for each noisy vector with different keys. Within each group, the SSIM metric was computed by comparing the similarity between each pair of images. To calculate the FID difference, we evaluate the generative images (for both watermarked and originally generative contents) on the MS-COCO 2014 training set for Stable Diffusion and on the ImageNet~\cite{deng2009imagenet} training set for the ImageNet diffusion model. We present the experimental results in Table \ref{tab:main_exp}. The results demonstrate that our method achieves comparable performance in both detection and identification tasks when compared to the post-hoc StegaStamp, achieving a tracing accuracy of 97.71\% and an AUC of 1 on average. We defer further discussion on the comparison with post-hoc methods to the \textit{Appendix}. In terms of image consistency, our method achieves an average of 0.998 SSIM between pairs of images with different watermarks, which significantly surpasses that of Tree-Ring. Besides, our method only imposes a negligible impact on the original generation quality by a slight increase on the original FID, which is comparable to the post-hoc watermark. Note that the Tree-Ring achieves nearly zero performance on the identification task. This might be attributed to the \textbf{imprecise latent inversion} and the \textbf{continuous watermarking space}, which makes watermarked examples less distinguishable when multiple users are engaged.

\begin{table*}[htbp]
\caption{This table reports WaDiff tracing accuracy (\%) and AUC under diverse data augmentations.}
\label{tab:2}
\setlength\tabcolsep{3pt}
\centering
    \begin{small}
    \begin{sc}
\resizebox{0.85\textwidth}{!}{\begin{tabular}{c|c|cccccccc}
\toprule
Model     &Case &Resize       & Blurring     & Color Jitter  & Noising & JPEG & Combine &Avg \\ \midrule
\multirow{3}{*}{\makecell[c]{Stable \\ Diffusion}} 
&AUC& 0.999  &0.999 &0.999 &0.997 &0.999  &0.999  &0.999  \\
& $\text{Trace}\ 10^4$ 
                    
                    & 97.02  & 97.14  & 96.00 & 88.52 &93.48 &93.02 & 94.19 \\ 
                    &$\text{Trace}\ 10^5$& 94.34   & 94.12 & 88.56 & 81.14 & 87.66&84.26& 88.34\\ 
                    &$\text{Trace}\ 10^6$& 89.46   &87.40&82.14  &72.50 & 80.30&78.04 & 81.64\\
                    
                    \midrule
\multirow{3}{*}{\makecell[c]{256$\times$256 \\ ImageNet}} 
&AUC& 0.999  &0.999 &0.999 &0.999 &0.999  &0.999  &0.999 \\
&$\text{Trace}\ 10^4$ & 98.90 & 94.48 &98.56  &91.80 &92.06&91.88& 94.61 \\ 
                    &$\text{Trace}\ 10^5$& 97.78 & 89.90 & 96.48 & 84.46 & 88.70&85.74 & 90.51\\ 
                    &$\text{Trace}\ 10^6$& 96.02  & 82.42  & 94.50 &76.26 &77.88&76.88 & 83.99 \\ 
                    \bottomrule
\end{tabular}
}
 \end{sc}
 \end{small}
\end{table*}

\subsection{Watermark Robustness Analysis}
\label{sec:5.3}
To evaluate our watermarking strategy against potential data augmentations, we adopt five commonly used augmentations including image resizing, image blurring, color jitter, Gaussian noising, and JPEG compression. Specifically, for resizing, we randomly resize the width and height of images within a range of 30\% to 80\% of their original size; for image blurring, we adopt the kernel size to 20; for color jitter, we randomly select from saturation (factor 1.5), contrast (factor 1.5) and sharpness (factor 1.5); for Gaussian noising, we add a Gaussian noise with $\sigma=0.1$; for the JPEG compression, we select the compression quality as 50. We also consider a combinational augmentation that incorporates image resizing, JPEG compression, and color jitter simultaneously. We report the identification performance under various augmentations in Table~\ref{tab:2}. The experimental results demonstrate the overall robustness of our method against diverse image augmentations, yielding an average tracing accuracy of 88.05\% for the Stable Diffusion model and 89.7\% for the ImageNet diffusion model. It is noteworthy that our method maintains an average AUC of 0.999 under various data augmentations and achieves identification performance of over 76\% among one million users even under combined data augmentation, which further validates the practicality and effectiveness of WaDiff. We provide robustness comparisons with other watermarking schemes in the \textit{Appendix}.

\begin{wraptable}{r}{6.9cm}
    \centering
   
    \caption{Detection results against different bit thresholds. $\text{P}_S$ and $\text{R}_S$ indicate the precision and recall for the Stable Diffusion respectively, where $\text{P}_I$ and $\text{R}_I$ represent the precision and recall for the ImageNet diffusion model respectively.}
    \label{tab:bit_threshold}
    \setlength\tabcolsep{1pt} 
    \begin{small}
    \begin{sc}
    \resizebox{0.57\textwidth}{!}{\begin{tabular}{c|cccccc}
        \toprule
        Metric    & $\tau_b=0.65$& $\tau_b=0.7$   & $\tau_b=0.75$  & $\tau_b=0.8$  &$\tau_b=0.85$ & AVG\\
        \midrule
          $\text{P}_S$& 0.982 & 0.997 & 1.000 &1.000&1.000 &0.996  \\
 
         $\text{R}_S$ & 0.998 & 0.994 & 0.978  &0.951&0.898 &0.964  \\

         $\text{P}_I$ & 0.989&0.998  &0.999 &1.000&1.000 & 0.997\\
         $\text{R}_I$ & 0.999& 0.999&0.997 &0.994 &0.983 &0.994  \\
        
        \bottomrule
    \end{tabular}
    }
    \end{sc}
\end{small}
\end{wraptable}

\subsection{Ablation Study}
\label{sec:5.4}
\paragraph{\textbf{Detection with Different Bit Thresholds.}} In addition to the reported metric AUC in our main results, in this section, we make a detailed analysis of our detection capability. In the detection task, the choice of the bit threshold $\tau_b$ would impact the overall detection performance. To investigate this, we vary the value of $\tau_b$ from 0.65 to 0.85 and examine the precision and recall of our detection performance. We conduct experiments on a dataset consisting of 5000 clean images and 5000 watermarked images, and the results are presented in Table~\ref{tab:bit_threshold}. Overall, our method demonstrates robustness to the selection of $\tau_b$, indicating its stability and reliability in detecting generative content.

\begin{figure}
\centering
\setlength{\belowcaptionskip}{-1cm}
\includegraphics[scale=0.27]{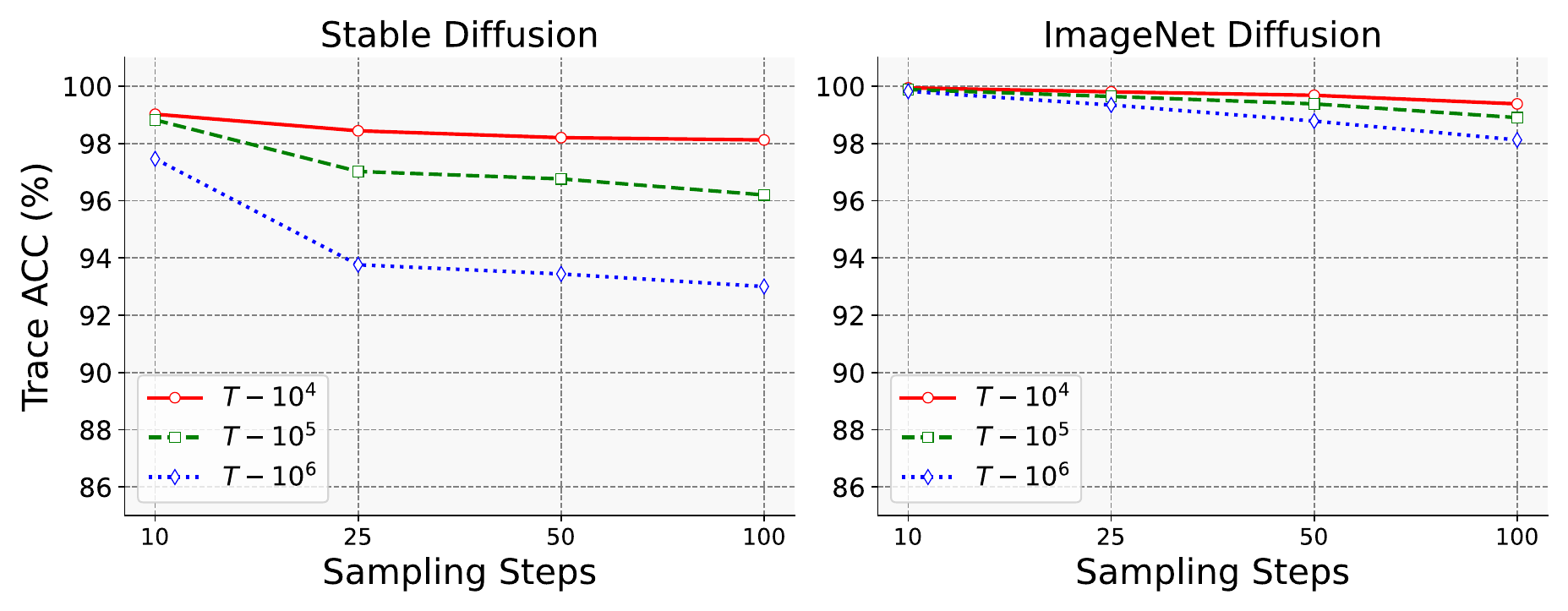}
\caption{The tracing accuracy results of two diffusion models with different DDIM sampling steps. We denote $T-m$ as tracing among $m$ users.}
\label{fig:4}
\end{figure}

\paragraph{\textbf{Experiments on Different Sampling Steps.}}
In this section, we analyze the impact of different sampling steps on our watermarking results. In addition to the 50 steps as default, we verify our method on 10, 25, and 100 DDIM sampling steps and demonstrate our experimental results in Figure \ref{fig:4}. The results demonstrate that our WaDiff achieves a stable performance across various sampling steps. Notably, it is observed that with a smaller sampling step, the tracing accuracy is boosted for both models, especially for the Stable Diffusion model, which shows an average increase of 2.3\% tracing accuracy for 10 sampling steps compared to the default 50 steps. This can be attributed to the fact that when the sampling step is small, the noisy image is not effectively recovered, allowing our watermark information to be more easily concealed in these flaw areas.

\begin{wraptable}{r}{6cm}
    \centering
    \caption{The tracing accuracy against two adaptive attacks, where DA and MA are short for the Diff and Multi-Message attack respectively.}
    \label{tab:adaptive_attack}
    \setlength\tabcolsep{1pt} 
    \begin{small}
    \begin{sc}
    \resizebox{0.5\textwidth}{!}{\begin{tabular}{c|ccc}
        \toprule
        Attack  &  Trace $10^4$ & Trace $10^5$ & Trace $10^6$\\
        \midrule
        DA (WaDiff)&77.50\% & 63.34\% & 49.42\%   \\
 
         DA (Stega) & 0.10\% & 0.00\% & 0.00\%   \\

         MA (WaDiff) & 97.96\% & 95.66\% & 91.48\% \\
         MA (Stega) &0.02\%  & 0.00\% &0.00\%  \\
        
        \bottomrule
    \end{tabular}
    }
    \end{sc}
\end{small}
\end{wraptable}

\subsection{Robustness Analysis with Adaptive Attacks}
\label{subsec:robustness}
In addition to the commonly used image augmentations discussed in Section \ref{sec:5.3}, we investigate two additional countermeasures to assess the robustness of our watermarking scheme. For the \textbf{Diff} Attack, we adopt a method similar to DiffPure~\cite{nie2022diffusion}, which uses a pre-trained diffusion model to purify our watermarked images. In our experiments, we set the number of diffusion and denoising steps to 20 and utilize the original DDPM sampler for this attack. In the \textbf{Multi-Message} Attack, we train a surrogate watermark encoder with the same architecture as in our experiments, but using different training data. We then embed a random watermark message into the watermarked example. Evaluation results against two adaptive attacks on both WaDiff and StegaStamp are presented in Table \ref{tab:adaptive_attack}. Our results demonstrate that WaDiff exhibits superior robustness compared to StegaStamp in these two adaptive attacks. This can be attributed to the integrated design of our watermarks, which are less vulnerable to diffusion denoising and are not easily removed by directly injecting post-hoc information.

\section{Conclusion and Limitations}

In this paper, we provide an efficient and robust watermarking framework WaDiff to not only detect whether an image is generated from our model but also identify the specific user who generated the image. WaDiff seamlessly incorporates a user-specific watermark as a conditioned input and applies fingerprinting to the generated contents during the image generation process. Extensive experiments demonstrate that WaDiff achieves accurate identification performance among a large number of users and remains robust under various data augmentations. While WaDiff still exhibits slightly inferior performance compared to the post-hoc watermarking method, our method fingerprints in the generation process which is more stealthy and hard to circumvent in practice. We hope that our work can pioneer the secure auditing of AI-generated content so that we can ensure that model usage aligns with regulatory requirements, compliance, and ethical standards, thereby enhancing the integrity of the generative models.



%
%
\bibliographystyle{splncs04}
\bibliography{main}

\clearpage
\appendix

\setcounter{footnote}{0}
\section{Implementation Details}
\label{sec:7}
In this section, we provide additional implementation details of our experiments. We implement our method based on the official code of the Stable Diffusion\footnote{https://github.com/CompVis/stable-diffusion} and the ImageNet diffusion model\footnote{https://github.com/openai/guided-diffusion}.
\subsection{Details of Pre-training Watermark Decoders}
To generate the training set for pre-training, we randomly select 5000 images from the training set of MS-COCO 2014 and resize the image size to $512\times512$ for the Stable Diffusion model; while for the ImageNet diffusion model, we randomly select 5000 images from the training set of ImageNet and resize the image size to $256\times256$ for training. For both cases, we set $\gamma$ to 20 and the epoch number to 300. To enhance the robustness of our watermark decoder, similar to the Stable Signature, we apply several data augmentations to the watermarked outputs before feeding them into the watermark decoder. These data augmentations include image blurring, color jitter, Gaussian noise, as well as JPEG compression. Notably, in our pre-training stage, our primary goal is not to guarantee good image quality for our watermark encoder, but rather to achieve a powerful watermark extraction capability within the watermark decoder. This is reasonable since, we discard the watermark encoder during the fine-tuning process, indicating that it does not affect the image quality of our watermarking framework.

\subsection{Details of Baseline Methods}
In this section, we introduce the implementation details for our baseline methods including the DwtDct, StegaStamp, and Tree-Ring.
\begin{itemize}
    \item \textbf{DwtDct} is a traditional post-hoc watermarking method that embeds fingerprinting information into the frequency domain. In our experiments, we follow a widely used implementation\footnote{https://github.com/ShieldMnt/invisible-watermark} and embed a 48-bit binary string to the image.

    \item \textbf{StegaStamp} is a deep-learning-based post-hoc watermarking strategy used in our pre-training stage. Similar to the training procedure in our pre-training stage, we adopt a 48-bit binary string as the watermarking message and slightly modify the hyperparameters by setting $\gamma$ to 15 and the number of epochs to 100 for both cases. Besides, to enhance the quality of watermarked images, we employ the perceptual loss instead of the original MSE loss, which improves the visual quality of images generated by the watermark encoder. Specifically, we simply utilize the first 23 layers of a pre-trained VGG-16 provided by PyTorch as our backbone for feature extraction and then calculate the MSE loss between the latent features of watermarked and original images as our perceptual loss.

    \item \textbf{$\text{Tree-Ring}_{Rand}$} Note that the authors of the Tree-Ring did not provide specific experimental settings for dealing with multiple-user cases. To compare the Tree-Ring method with our approach, we extend it to the multiple-user scenario without modifying its original watermarking setting. Specifically, in the $\text{Tree-Ring}_{Rand}$ case, we randomly sample a unique Gaussian noise pattern assigned to each user. We replace the original pixels in the Fourier domain with the generated key pattern. For owner identification, we reverse the candidate image into the initial noisy vector and then determine the owner by finding the closest match to the Gaussian pattern.
    
    \item \textbf{$\text{Tree-Ring}_{Rings}$} For the $\text{Tree-Ring}_{Rings}$ case, we use the same number of rings, which is 10, to generate the key pattern. We follow the original watermarking strategy and assign each ring a random value sampled from a Gaussian distribution to differentiate between users. Similar to the $\text{Tree-Ring}_{Rand}$ case, we conduct owner identification by identifying the user with the closest key pattern.
\end{itemize}

\subsection{Message Retrieval Loss for Stable Diffusion}
Note that in Stable Diffusion, the image is compressed by a latent encoder into a latent feature, and the diffusion process occurs in the feature space rather than the pixel space. Therefore, directly applying Equation \ref{eq4} would only yield the reversed feature. To calculate the message retrieval loss, we need to decode the reversed feature through the latent decoder to obtain the single-step-reverse image. We then optimize the message retrieval loss according to Equation \ref{eq5}. During the fine-tuning process, both the latent encoder and decoder remain fixed.

\subsection{Consistency Loss for ImageNet Diffusion Model}
In this section, we provide more details on the implementation of the consistency loss for the ImageNet diffusion. Instead of directly aligning the output of our WaDiff model with the original output, we take an additional step by applying Equation \ref{eq4} to both outputs. This gives us the single-step-reverse image $\mathbf{x}_{0, i}^t$ for our WaDiff model and $\mathbf{x}_0^t$ for the original model. To measure the consistency between these images, we adopt the perceptual loss by replacing the consistency loss in Equation \ref{eq6} with $\mathcal{L}_c(b(\mathbf{x}_{0, i}^t), b(\mathbf{x}_0^t))$, where $b$ is a well-trained backbone for feature extraction. Similarly, in our approach, we use the first 23 layers of a pre-trained VGG-16 as $b$ which is the same for training the StageStamp. 
\section{Additional Ablation Studies}
In this section, we present additional ablation studies. We provide the tracing accuracy results along with the SSIM and FID differences. It is worth noting that in order to present the SSIM and FID difference values on the same scale, we apply min-max normalization to both the SSIM and FID difference values before presenting the results.
\subsection{Different Fine-tuning Sections}
\begin{figure}
\centering
\includegraphics[scale=0.27]{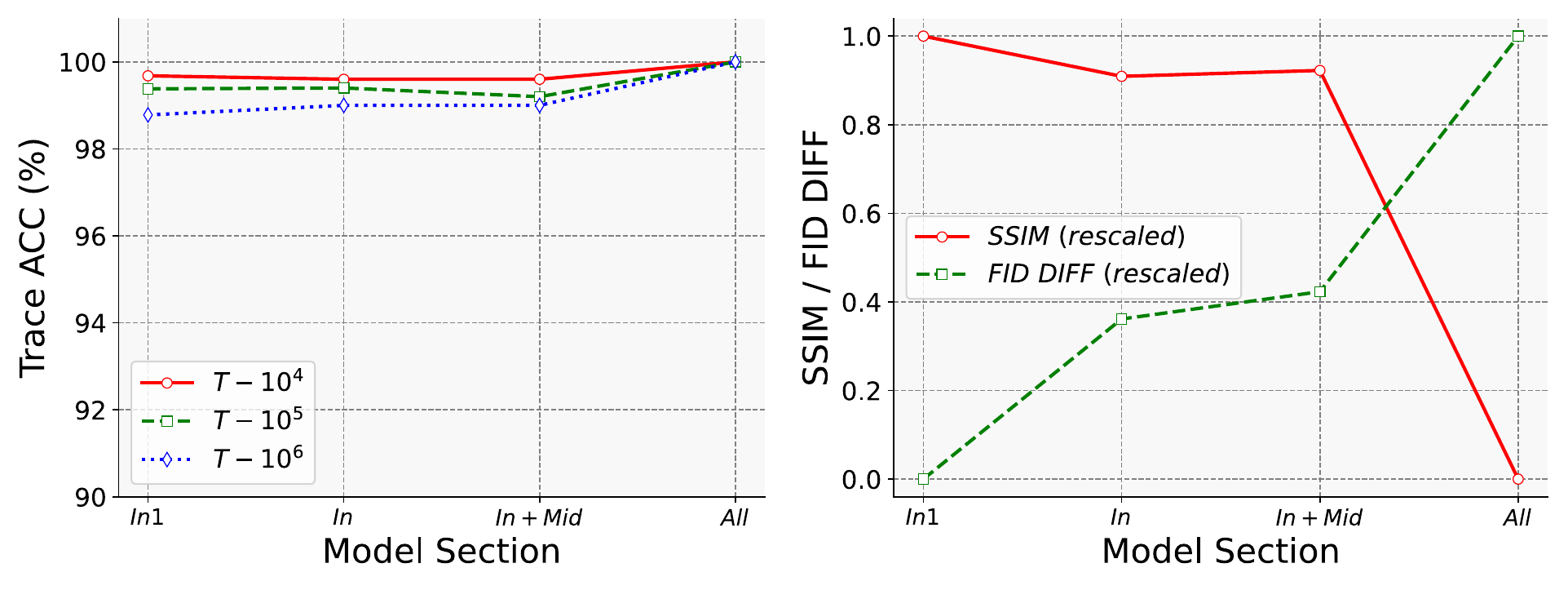}
\caption{The tracing accuracy, along with the SSIM and FID difference, are demonstrated against different fine-tuning sections. Note that both the SSIM and FID difference values have been rescaled using the min-max algorithm to a range of $[0, 1]$.}
\label{fig:5}
\end{figure}

We performed fine-tuning on various sections of the ImageNet diffusion model. Specifically, in addition to fine-tuning the first input block (referred to as $In1$), we also fine-tuned the entire input blocks ($In$), the entire input and middle blocks ($In + Mid$), and the entire architecture ($All$). The experimental results are presented in Figure \ref{fig:5}. The results indicate that fine-tuning only the first block is sufficient to achieve comparable tracing performance compared to fine-tuning additional model sections. Additionally, we observed that fine-tuning the entire model architecture led to a decrease in both the generation quality and the Structural Similarity Index (SSIM). Therefore, it is evident that fine-tuning the entire architecture may not be desirable due to the trade-off between tracing performance and generation quality.

\subsection{Different Bit Lengths}
\begin{figure}
\centering
\includegraphics[scale=0.27]{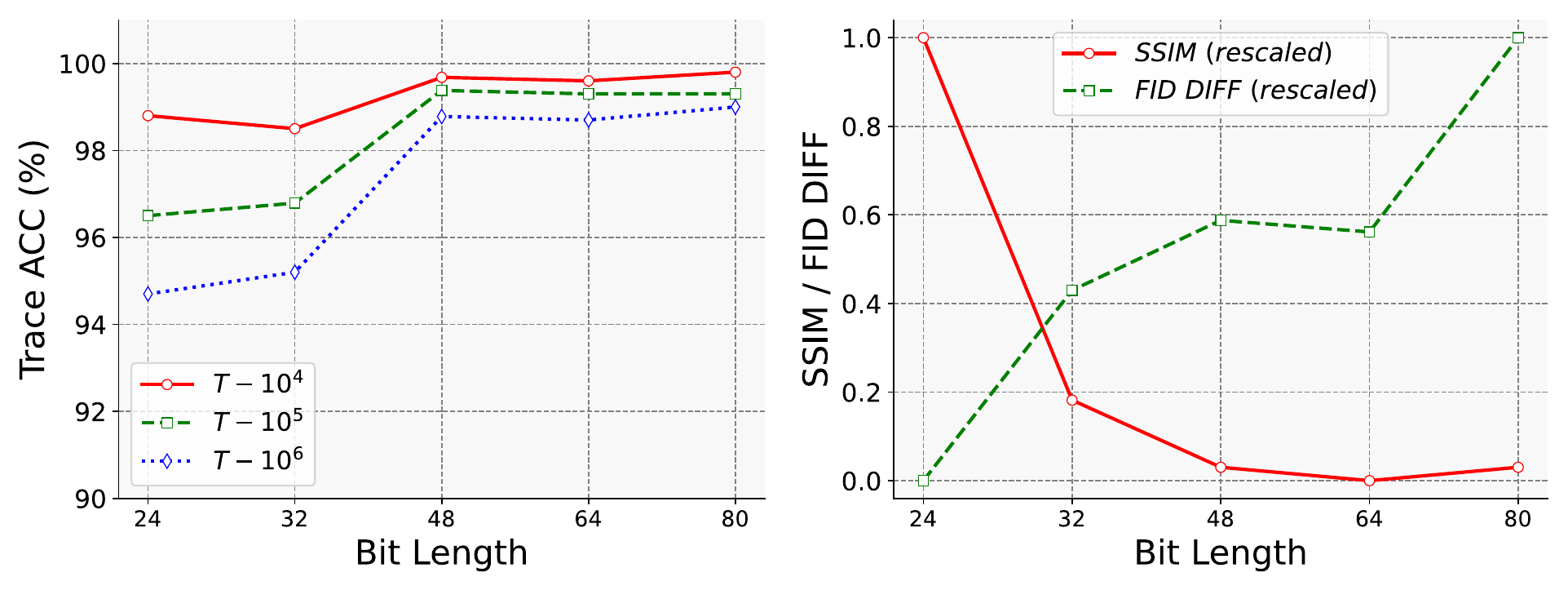}
\caption{The tracing accuracy, along with the SSIM and FID difference, are demonstrated against different bit lengths. Note that both the SSIM and FID difference values have been rescaled using the min-max algorithm to a range of $[0, 1]$.}
\label{fig:6}
\end{figure}

In this section, we present additional experimental results on watermarks with varying bit lengths. We conducted experiments with bit lengths ranging from 24 to 80 and analyzed the results using the ImageNet diffusion model, as depicted in Figure \ref{fig:6}. Our findings demonstrate that WaDiff maintains robust tracing performance even when the watermarking budget is reduced by half, from 48 bits to 24 bits. In the case of one million users, this reduction only leads to a marginal drop of approximately 4\% in tracing accuracy. Moreover, in practical scenarios, one can enhance the tracing robustness by simply increasing the number of watermark bits, since there is not much computational increase in our method.

\subsection{Different Time Thresholds}
\begin{figure}
\centering
\includegraphics[scale=0.27]{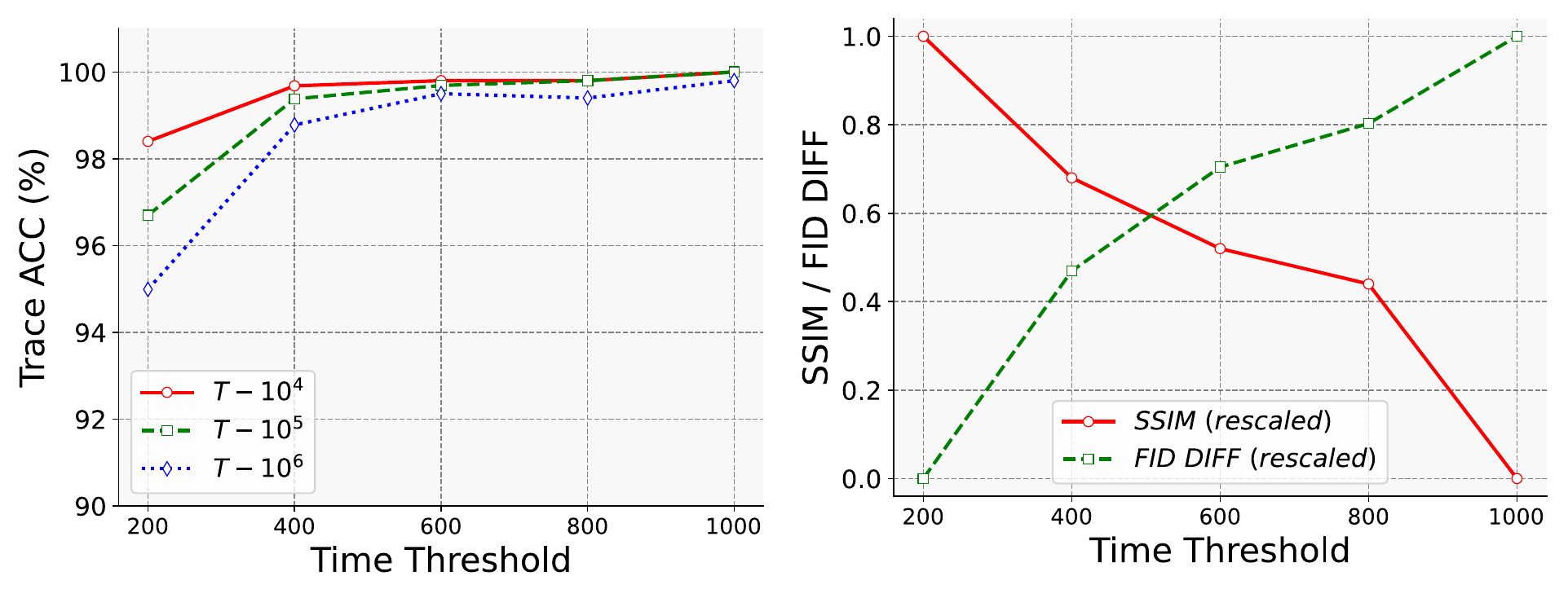}
\caption{The tracing accuracy, along with the SSIM and FID difference, are demonstrated against different time thresholds. Note that both the SSIM and FID difference values have been rescaled using the min-max algorithm to a range of $[0, 1]$.}
\label{fig:7}
\end{figure}

We conduct fine-tuning on the ImageNet diffusion model using different values of $\tau$ (ranging from 200 to 1000). We present our experimental results in Figure \ref{fig:7}. The results reveal that as $\tau$ is decreased, the tracing accuracy also decreases. Although increasing $\tau$ can improve the tracing accuracy, we still observe a decrease in the SSIM. This observation is reasonable because a higher $\tau$ implies that the null watermark is applied over fewer time steps, leading to reduced similarity between samples with different watermarks.

\subsection{Different $\eta$}
\begin{figure}
\centering
\includegraphics[scale=0.27]{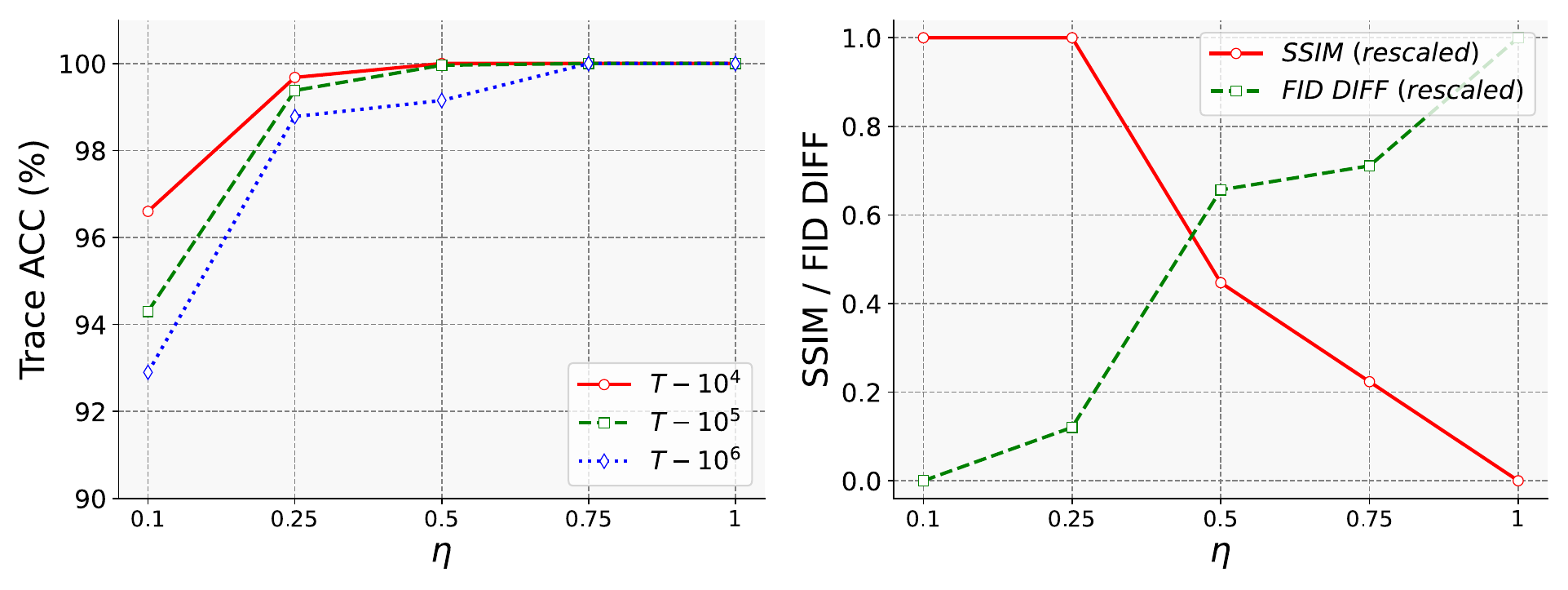}
\caption{The tracing accuracy, along with the SSIM and FID difference, are demonstrated against different $\eta$. Note that both the SSIM and FID difference values have been rescaled using the min-max algorithm to a range of $[0, 1]$.}
\label{fig:8}
\end{figure}

In this section, we perform experiments with different values of $\eta$ ranging from 0.1 to 1 using the ImageNet diffusion model and present the results in Figure \ref{fig:8}. We observe that a smaller value of $\eta$ would decrease the tracing accuracy. Although increasing $\eta$ leads to an improved tracing accuracy, it yields degradation in generation quality as indicated by an increased FID difference value.

\subsection{Different Guidance Scales}
During the generation process of Stable Diffusion, users have the option to manually select a guidance scale, enabling them to control the balance between generation specificity and diversity. A higher guidance scale corresponds to images that closely align with the provided text prompt but with lower generation diversity. To evaluate the potential impact of the guidance scale, we conducted experiments on the Stable Diffusion using various scales ranging from 5 to 20. The experimental results, as shown in Table~\ref{tab:3}, demonstrate our watermarking strategy is effective across different guidance scales. We achieved an average tracing accuracy of 97.56\% across $10^4$ users, 95.43\% across $10^5$ users, 91.82\% across $10^6$ users, and a stable AUC of 0.999, which indicates that the guidance scale has a negligible impact on the effectiveness of our method.

\begin{table}[t]
    \caption{Tracing accuracy (\%) against different guidance scales.}
    \label{tab:3}
    \setlength\tabcolsep{3pt} 
    \begin{center}
    \begin{small}
    \begin{sc}
    \resizebox{0.8\textwidth}{!}{\begin{tabular}{c|ccccc}
        \toprule
        \makecell[c]{Guidance \\ Scale}    & $\text{Trace}\ 10^4$   & $\text{Trace}\ 10^5$  & $\text{Trace}\ 10^6$  &Trace Avg & AUC\\
        \midrule
          5& 97.62 & 94.20 & 90.40 &94.07&0.999   \\
 
         7.5 & 98.20 & 96.76 & 93.44  &96.13&0.999   \\

         10 & 97.78&96.66  &93.94 &96.13&0.999  \\
         15 & 97.28& 95.94&92.02 &95.08 &0.999   \\
         20 &96.92 &93.60 &89.28 & 93.27&0.999   \\
        \bottomrule
    \end{tabular}
    }
    \end{sc}
\end{small}
\end{center}
\end{table}

\subsection{Different Schedulers}
We also conducted experiments using different schedulers for image generation, including both the PLMS and DPMSolver. These experiments were performed on the Stable Diffusion model using the official implementation available in the original code repository. For the DPMSolver, we utilized 15 sampling steps due to its high sampling efficiency. Our observations reveal that our method consistently achieves high tracing accuracy across various schedulers. Specifically, with the PLMS scheduler, we obtained a tracing accuracy of 98.20\%, 96.00\%, and 93.26\% for $10^4$, $10^5$, and $10^6$ users respectively. With the DPMSolver scheduler, we achieved a tracing accuracy of 97.98\%, 95.72\%, and 92.40\% for $10^4$, $10^5$, and $10^6$ users respectively. These results highlight the robustness and effectiveness of our method across different scheduling approaches.

\begin{figure*}
\centering
\includegraphics[scale=0.27]{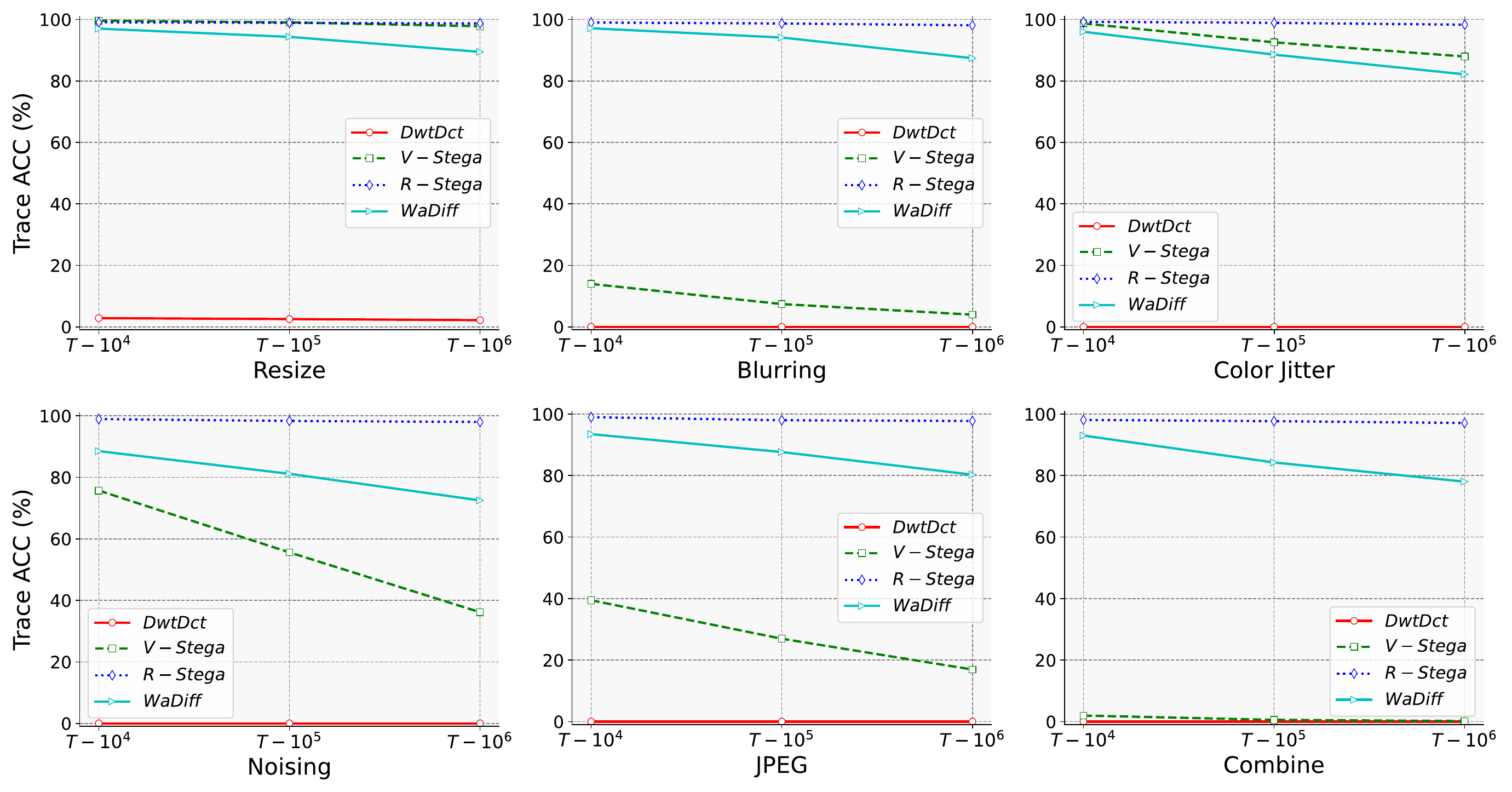}
\caption{Robustness analysis of the Stable Diffusion, where V-Stega indicates the vanilla StegaStamp, and R-Stega represents the robust variant incorporating data augmentations during training.}
\label{fig:sd_rob}
\end{figure*}

\section{Additional Robustness Analysis}
\begin{figure*}
\centering
\includegraphics[scale=0.27]{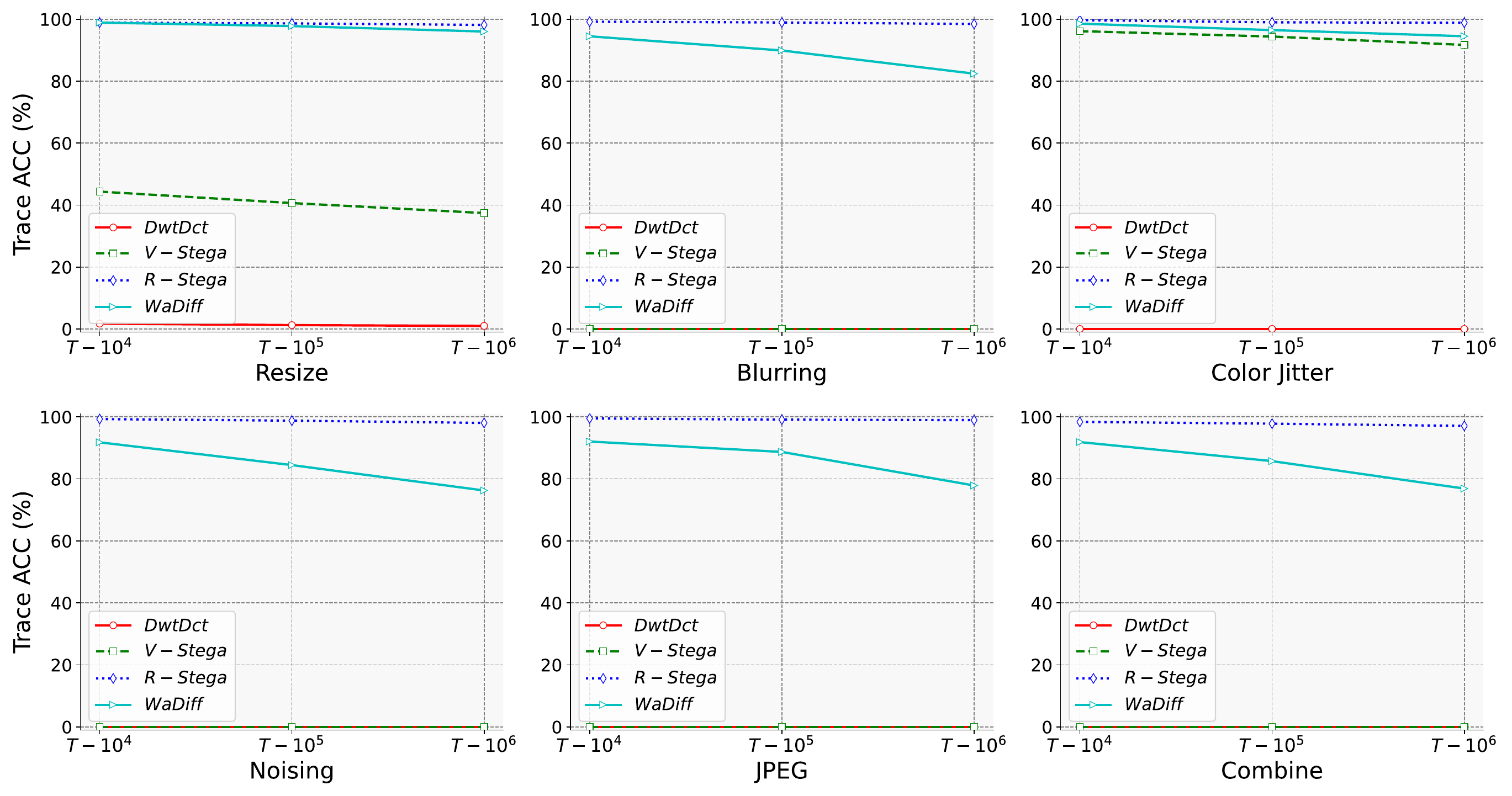}
\caption{Robustness analysis of the ImageNet diffusion model, where V-Stega indicates the vanilla StegaStamp, and R-Stega represents the robust variant incorporating data augmentations during training.}
\label{fig:image_rob}
\end{figure*}

In this section, we present additional comparisons of WaDiff with DwtDct and StegaStamp, considering various data augmentation techniques. For StegaStamp, we utilize two different training schemes: a vanilla StegaStamp and a robust variant that incorporates data augmentations during the training phase. Both schemes employ the perceptual loss. We also extended the total number of training epochs to 300 for the robust StegaStamp, as the inclusion of data augmentations requires more time to achieve convergence. The experimental results for both the Stable Diffusion and ImageNet diffusion model are shown in Figure \ref{fig:sd_rob} and \ref{fig:image_rob}, respectively. Our experiments demonstrate that WaDiff significantly outperforms the vanilla StegaStamp on most augmentations. Although the robust StegaStamp achieves slightly higher performance, it comes at the cost of significant degradation in image quality, which can be easily detected through human inspections.

\section{Comparisons with Post-hoc Watermark Schemes}

We emphasize that the post-hoc methods provided are solely intended for reference purposes, and direct comparisons between the identification performance of WaDiff and post-hoc methods are inappropriate due to their \textbf{different threat models}. Post-hoc methods implant watermarks into the ground-truth images, which significantly simplifies the fingerprinting process. In contrast, our approach integrates fingerprinting into the generation process, resulting in improved stealthiness and efficiency. Consequently, even with white-box access, our watermarking framework poses significant challenges for direct attacks. Furthermore, while post-hoc watermarks may exhibit good identification accuracy, they are susceptible to potential vulnerabilities, as discussed in Section~\ref{subsec:robustness}. Additionally, their separable characteristics make them easily removable in the open-source stable diffusion model by simply commenting out a single line of code.

\section{Additional Watermarked Examples}
\begin{figure*}
\centering
\includegraphics[width=12cm, height = 17cm]{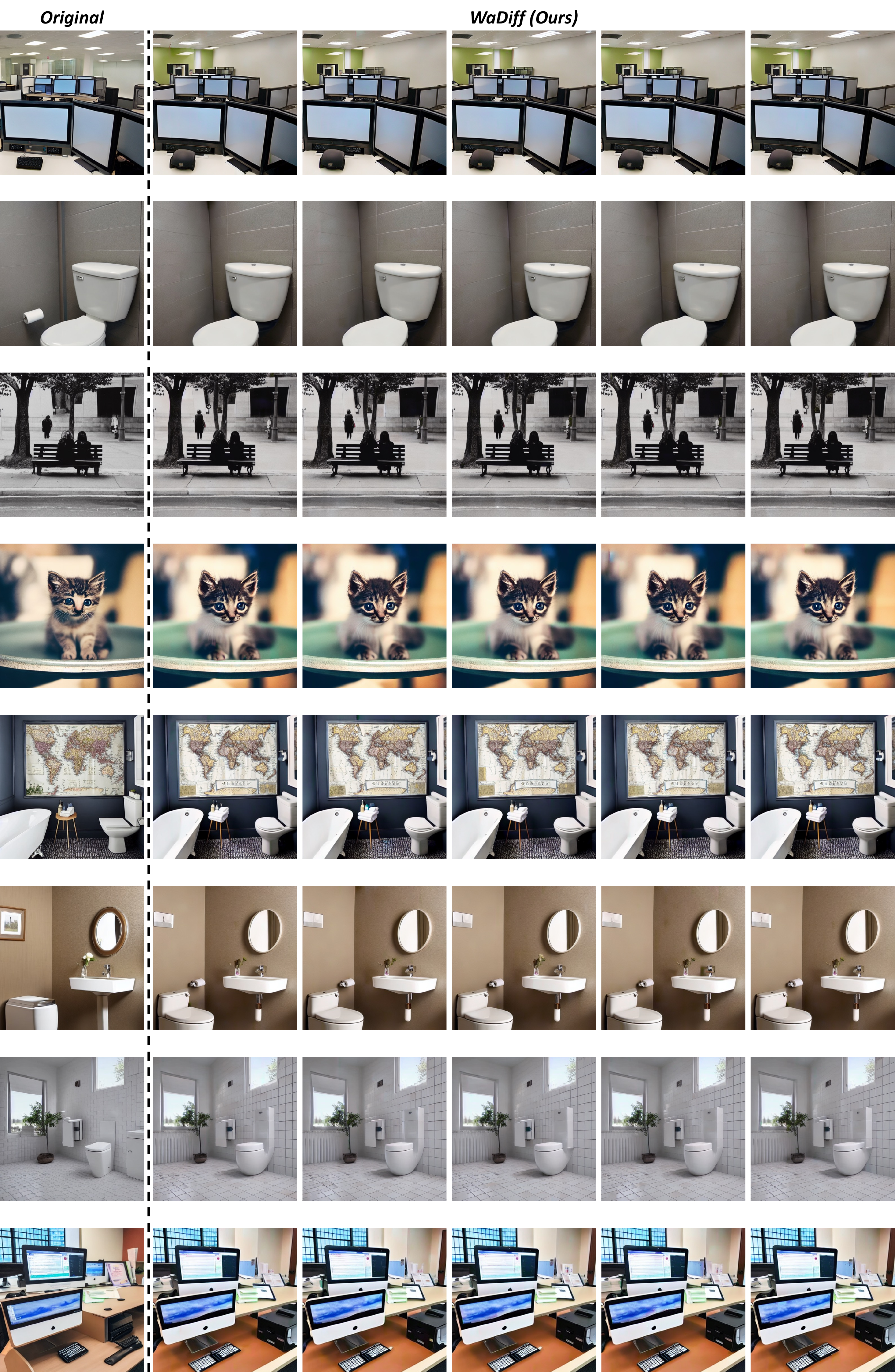}
\caption{We demonstrate additional original images along with the watermarked images generated by our WaDiff. All the images are sampled from the Stable DIffusion with the text prompts from the MS-COCO 2014 validation set.}
\label{fig:coco_all}
\end{figure*}

\begin{figure*}
\centering
\includegraphics[width=12cm, height = 17cm]{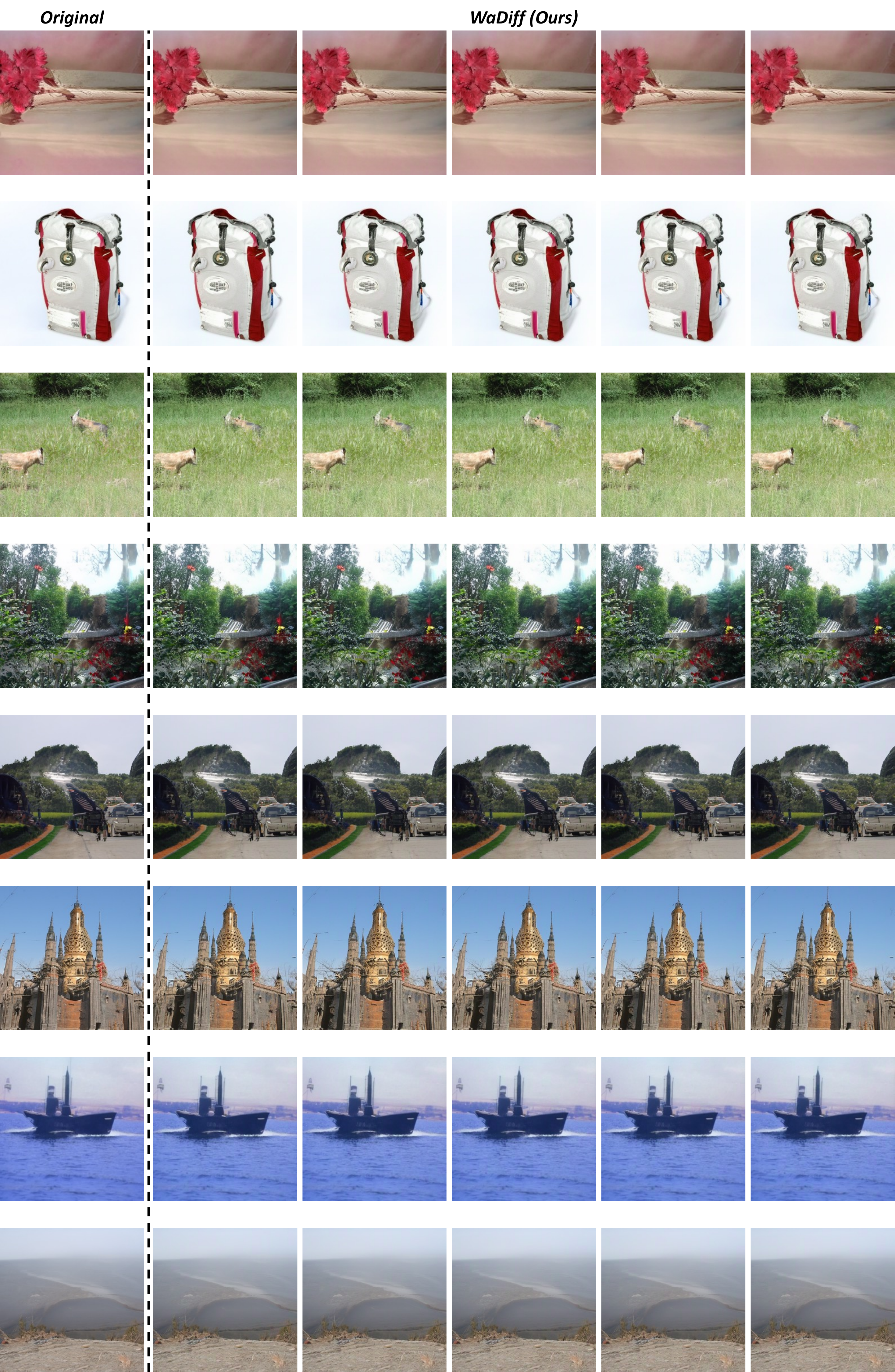}
\caption{We demonstrate additional original images along with the watermarked images generated by our WaDiff. All the images are sampled from the 256 $\times$ 256 ImageNet diffusion model.}
\label{fig:imagenet_all}
\end{figure*}

We provide additional watermarked examples for the Stable Diffusion in Figure \ref{fig:coco_all} and the ImageNet diffusion model in Figure \ref{fig:imagenet_all}.

\end{document}